\documentclass[prl,showpacs,amsmath,amssymb,twocolumn]{revtex4}

\newcommand{\be}{\begin{equation}}
\newcommand{\ee}{\end{equation}}
\newcommand{\la}{\label}
\newcommand{\dd}{{\rm{d}}}
\newcommand{\qq}{\mathcal{Q}}

\usepackage{graphicx}
\usepackage{amsmath,bm}

\begin{document}

\title{Non-equilibrium thermodynamics for functionals of current and density}

\author {Vladimir Y. \surname{Chernyak}$^a$}
\email{chertkov@lanl.gov,   chernyak@chem.wayne.edu, malinin@chem.wayne.edu,  razvan@lanl.gov}
\author{Michael \surname{Chertkov}$^b$}
\author{Sergey V.  \surname{Malinin}$^a$}
\author{Razvan  \surname{Teodorescu}$^b$}
\affiliation{$^a$Department of Chemistry, Wayne State University, 5101 Cass Ave,Detroit, MI 48202}
\affiliation{$^b$Theoretical Division and Center for Nonlinear Studies, LANL, Los Alamos, NM  87545}
\date{\today}

\begin{abstract}
We study a stochastic many-body system maintained in an
non-equilibrium steady state. Probability distribution functional of
the time-integrated current and density  is shown to attain a
large-deviation form in the long-time asymptotics. The corresponding
Current-Density Cram\'{e}r Functional (CDCF) is explicitly derived
for irreversible Langevin dynamics and discrete-space Markov chains.
We also show that the Cram\'{e}r functionals of other linear
functionals of density and current, like work generated by a force,
are related to CDCF in a way reminiscent of variational relations
between different thermodynamic potentials. The general formalism is
illustrated with a model example.
\end{abstract}


\maketitle

\paragraph{1. Introduction}
Deriving probability distributions of relevant physical quantities
for stochastic systems out of equilibrium is a complex problem of
great interest for many experimental situations. The issue is
difficult because of the complexity of non-equilibrium evolutions.
In recent years, several models were proposed where such quantities
could be computed. For example, the case of steady-state,
non-equilibrium systems  was studied in \cite{BSGJL}, using
theoretical tools developed earlier \cite{DV, BADZ}.

In the case when the stochastic system can be described by equations
of hydrodynamic type, a special matrix-based  formulation can be
used \cite{DEHP}, leading to a rich class of behaviors, including
integrability \cite{Schutz}, non-ergodicity and non-equilibrium
phase transitions \cite{RPS}, and interesting connections to random
matrix theory \cite{RMT}.

In the present work, we investigate generic stochastic systems out
of equilibrium, in the large deviation limit, under the assumption
of continuous symmetries for the space of trajectories. This allows
the large deviation functional for density and currents to be
derived, using a different approach than in \cite{BSGJL, RPS}.

\paragraph{2. Langevin processes in symmetric spaces}

Consider a Langevin process given in a configuration space $M$ by a stochastic equation 
 \be 
 \la{bare} 
 \dot{\eta}_i = F_i(\bm{\eta}) +\xi_i(t), 
 \quad \langle \xi_i(t) \xi_j(t') \rangle =  \delta_{ij}\delta(t-t'), 
 \ee
or equivalently by a path integral with a measure 
 \be 
 \la{q} 
 \dd \qq =Z^{-1}\exp\left \{ -S [ {\bm \eta}(\tau)] \right \} {\cal D}{\bm \eta}(\tau), 
 \ee 
for stochastic paths
(maps $\mathbb{R}_+ \to M$) $\bm{\eta}(t)$, with $\mathcal{D} \bm \eta$ , $Z$, and
\begin{eqnarray}
\label{action} S( {\bm \eta}) \!=\! (1/2)\int_{0}^{t}d{\tau} \,
\left[ \dot{\eta}_{i}-F_{i}({\bm
\eta})\right] \left[ \dot{\eta}_{i}-F_{i}({\bm \eta})\right ]
\end{eqnarray}
being the functional measure, a normalization factor, and the action, respectively. Hereafter,
summation over repeated indices is assumed. We use rescaled units so that the temperature of the
thermal bath, modeled by the Langevin term in (\ref{bare}), is equal to one. We are interested in
large deviation ($t\to\infty$) relations for stochastic trajectories of the process (\ref{bare}),
in the case when the system exhibits continuous symmetries preserving the metric induced by the
force field $\bm F$ (isometries).  For a compact representation of stochastic trajectories, we
introduce the fluctuating density and current at a point ${\bm x}$ of the trajectory's 
configuration space
\begin{eqnarray}
 & \varrho ( \bm{\eta} ,\bm x )   \equiv  \frac{1}{t}\int_0^t d \tau \delta(\bm{x} - \bm{\eta}(\tau)),   
 \la{rho-jay1} \\
 & \gimel_i ( \bm{\eta}, \bm x )  \equiv  \frac{1}{t}\int_0^t d \tau \dot{\eta}_i
\delta(\bm{x} - \bm{\eta}(\tau)).   \la{rho-jay2}
\end{eqnarray}
In the next section we will derive the large-deviation limit of the
joint probability distribution function for $\varrho, {\bm \gimel}$:
\be 
\la{prob} 
\mathcal{P}[ \bm{J}({\bm x}) , \rho({\bm x})] 
\equiv \langle \delta(\varrho-\rho) \delta(\bm{\gimel} -
\bm{J})\rangle_{\xi} 
\,.
\ee 
We will show that in the $t\to\infty$
limit, it takes the form $\mathcal{P}(\bm{J}, \rho) \sim
\exp [-t\mathcal{S}(\bm{J}, \rho)]$, where the CDCF is
\begin{align}
\label{CDDF-continuous-explcit} {\cal S}({\bm
J}, \rho )=\int_{M}d{\bm x} \frac{
[ \bm{F}\rho-\bm{J}-(1/2)\bm{\partial}\rho]^2}{2\rho} .
\end{align}
We will also obtain a similar large-deviation result for discrete
Markov processes.

The explicit large deviation result (\ref{CDDF-continuous-explcit}) can be immediately used to get
thermodynamics-like relations for Cr\'{a}mer functions of derived objects. One introduces the
vector and scalar potentials (which can also be interpreted as gauge fields, i.e. generators of
continuous symmetry transformations mentioned above), $V({\bm x}), \bm{A}({\bm x})$ and the
corresponding charges
\begin{eqnarray}
 \label{define-J-rho} & w_{{\bm A}}(\bm{\gimel})\equiv\int_{M}d{\bm x}A_{j}({\bm x})
 \gimel^{j}({\bm\eta};{\bm
x})=\frac{1}{t}\int_{0}^t d \tau \dot{\eta}^{j}A_{j}({\bm \eta}),
\nonumber \\ & u_{V}(\varrho)\equiv\int_{M}d{\bm x}V({\bm
x})\varrho({\bm\eta};{\bm x})=\frac{1}{t}\int_{0}^{t}d\tau
V({\bm\eta}(\tau)).
\end{eqnarray}
At $t\to\infty$, the joint p.d.f. $\mathcal{P}(w,u)\equiv \langle \delta(w-w_{{\bm
A}}(\bm{\gimel})) \delta(u - u_{V}(\varrho))\rangle_{\xi}$ of $w_{{\bm A}}(\bm{\gimel})$ and
$u_{V}(\varrho)$ has the large deviation form 
 $\mathcal{P}_{[{\bm A}, V ]} (w,u)\sim \exp(-t{\cal S}_{[{\bm A},V]}(w, u))$, 
where
\begin{align}
\label{variational-pinciple} & {\cal S}_{[ {\bm A}, V]}(w,
u)=\inf_{w_{{\bm A}}({\bm J})=w, \; u_{V}(\rho)=u} {\cal S}( {\bm
J},\rho).
\end{align}

\paragraph{3. Derivation for the Langevin processes}
Using a standard representation for the Dirac $\delta$-functional
(\ref{prob}) gives for the
probability $\mathcal{P}(\bm{J}, \rho)\sim$
 \begin{align} \nonumber
 & 
 \int\!\!{\cal D}{\bm A}({\bm x}) {\cal D}V({\bm x})
e^{-it\int d{\bm x}({\bm A}{\bm J}+V\rho)}\!\!
 \int\!\! {\cal D}{\bm{\eta}}(t) e^{-S({\bm\eta};{\bm A},V)}, \\
 \label{S-modified}
 & S({\bm\eta};{\bm A},V)=S({\bm\eta})-i \int_{0}^{t} d \tau
 \left( \dot{\eta}^{j}A_{j}({\bm \eta})+ V({\bm\eta})\right),
 \end{align}
where ${\cal D}{\bm A}({\bm x})$, ${\cal D}V({\bm
x})$ are the standard field-theoretical notations for functional
differentials/measures.

In the large deviation limit ($t\to\infty$),  the path
integral in (\ref{S-modified}) is estimated as
 \begin{align}
 & \int\!\! {\cal D}{\bm{\eta}}(\tau) e^{-S({\bm\eta};{\bm A},V)}
 \propto\exp\left(t\mathcal{F}(\bm{A}, V)\right), \mbox{ where}
 \label{FAV}\\
 & \mathcal{F}(\bm{A}, V)=\int d{\bm x}
 \hat{{\cal L}}_{{\bm A},V}\bar{\rho}({\bm x}), \mbox{ and}
 \label{braket}\\
 &  \hat{{\cal L}}_{{\bm A},V}\!\equiv\!(1/2)\nabla_j\nabla_j\!-\!\nabla_{j}F_{j}\!+\!iV,
 \  \nabla_j \!\equiv\!\partial_j - i
 A_j, \label{FP-modified}
 \end{align} 
with $\bar{\rho}({\bm x})$ the normalized right-eigenvector of
the ground state (lowest eigenvalue) for the Fokker-Planck operator
$\hat{{\cal L}}_{{\bm A},V}$,
\begin{eqnarray}
\hat{{\cal L}}_{{\bm A},V}\bar{\rho}=\lambda\bar{\rho}, \mbox{ and }
\int d{\bm x}\bar{\rho}({\bm x})=1.
\label{lambda}
\end{eqnarray}

Applying further the saddle-point approximation for the functional integral in
(\ref{S-modified}) with respect to ${\bm A},V$, and using
(\ref{S-modified},\ref{FAV},\ref{braket},\ref{FP-modified}), we arrive at the following equations:
\begin{eqnarray}
 {\bm J}(x)=({\bm F}+i{\bm A}-(1/2)\bm\partial)\bar{\rho}({\bm x}),\ \ \bar{\rho}=\rho.
 \label{sp}
\end{eqnarray}
Solving (\ref{lambda},\ref{sp}) for ${\bm A},V$ and $\bar{\rho}$ and substituting the result back
in the saddle expression for (\ref{S-modified}) yields (\ref{CDDF-continuous-explcit}). Note that
the final result does not depend on $\lambda$, although it explicitly enters
(\ref{lambda}).

Notice that for the physical current, $\partial_i J_i=0$, and the functional integration measure in
(\ref{S-modified}) is invariant with respect to the gauge transformation, $A_j\to
A_j+\partial_j\varphi$. In deriving (\ref{FAV},\ref{braket}) from (\ref{S-modified}), the gauge
freedom was fixed by the requirement that the left eigenfunction of $\hat{\cal L}_{{\bm A},V}$
conjugated to the right eigenfunction, $\bar{\rho}$, equals unity.

\paragraph{4. Comments}
(i) The variational principle (\ref{variational-pinciple})
establishes the optimal-fluctuation picture of the distributions
${\cal P}(w, u;t)$ at long times on the level of currents/densities:
The relevant probability ${\cal P}(w, u ;t)$ is determined by the
probability ${\cal P}({\bm J},\rho;t)$ of the optimal distributions
$({\bm J}({\bm x}),\rho({\bm x}))$ of current and density that
correspond to the maximal probability, provided they reproduce the
correct set  of rates $(w, u)$ as in (\ref{variational-pinciple}).

(ii) Comparison of the probability of a stochastic trajectory
${\bm\eta}$ and its time-reversed counterpart, combined with the
fact that the entropy production along ${\bm\eta}$ depends on the
correspondent current distribution ${\bm \gimel}(\{\bm{\eta}\}, \bm x)$
only, leads to the fluctuation theorem relation \cite{FT} for the
CDCF
\begin{align}
 \label{FT-for-CDF} 
 & {\cal S}(-{\bm J},\rho)-{\cal S}({\bm J},\rho)=
 2w_{{\bm F}}({\bm J}).
\end{align}

(iii)  The special choice of the gauge (\ref{sp}) has
a simple geometric interpretation: solving a heat-kernel type
equation for operator (\ref{FP-modified}) with delta-function
initial condition usually leads to a gaussian solution with
increasing variance. However, it is possible to apply infinitesimal
gauge transformations such that the solution remain singular at all
times. The gauge (\ref{sp}) achieves  this result.

(iv) The fact that in the large time limit $t \to \infty$ it is possible to approximate the
effective action with a quadratic form in the gauge fields (which is equivalent to
(\ref{CDDF-continuous-explcit})), can be traced back to the formal expansion of the (gauged) heat
kernel \cite{Seeley} \be \langle x | e^{tD} | x\rangle  = (4 \pi t)^{-d/2} \left [ 1 + t a_1 +
\ldots \right ], \ee where $D = \nabla^{\dag}_\mu \nabla^{\mu} + X $ is a gauged heat kernel in $d$
dimensions, and $a_1$ is the first Seeley coefficient, $a_1 = - X - \omega(A_{\mu})$, with $\omega$
a quadratic form. This result holds for compact spaces; for a generalization to spaces with
(possibly fractal) boundary, see \cite{Duplantier}.

(v) The result (\ref{CDDF-continuous-explcit}) can also be understood in very general terms
starting with the problem of computing the current algebra (and hamiltonian) of sigma models in
symmetric spaces (a field theory for maps into Riemann spaces with simple compact isometry group).
In this set-up, the force field $\bm F$ provides a metric, and the continuous symmetry group
preserves the metric. It is known \cite{Abdalla} that the functional of currents for this theory
has the form (\ref{CDDF-continuous-explcit}). This result was also used in the context of
mesoscopic transport theory \cite{Levitov}.

\paragraph{5. Derivation for irreversible Markov chains}
Irreversible Markov Chain (MC) stochastic dynamics can be viewed as
a discrete (regularized) counterpart of the Langevin processes, that
also occur in continuous time. We start with formulating the MC
dynamics such that the formalism reported above can be applied  with
only minor changes. The configuration space $M$ is a graph that
consists of vertices $a\in M_{0}$ and edges $\{a,b\}\in M_{1}$.  An
edge $\{ a, b \}$ will be further denoted $a\rightarrow  b$ or
$b\leftarrow a$. A Markov process is determined by a set of rates
$k_{ab}\ne k_{ba}$ that reside on the graph edges. A discretized
stochastic trajectory
${\bm\eta}=(a_{n},\ldots,a_{0};\tau_{n},\ldots,\tau_{1})$ with
$a_{j+1}\rightarrow  a_{j}$ and $0<\tau_{1}<\ldots<\tau_{n}<t$
represents a set of instantaneous jumps from $a_{j-1}$ to $a_{j}$
that occur at times $\tau_{j}$ respectively. The Markovian measure
${\cal D}{\bm\eta}\exp\left(-S({\bm\eta})\right)$, with ${\cal
D}{\bm\eta}=\sum_{n}d\tau_{1}\ldots d\tau_{n}$, is described by the
action
\begin{align}
\label{action-discrete} &
\exp\left(-S({\bm\eta})\right)=
\prod_{j=1}^{n}k_{a_{j}a_{j-1}}e^{-\sum_{i=0}^{n}\kappa_{a_{i}}(\tau_{i+1}-\tau_{i})}\,,
\end{align}
where $\kappa_{a}=\sum_{b\rightarrow  a}k_{ba}$ is the rate of leaving the site $a$.

Similar to (\ref{rho-jay1},\ref{rho-jay2}) , we can define the density $\rho_{a}({\bm\eta})$ and the current
$J_{ab}({\bm\eta})$ associated with a trajectory ${\bm\eta}$, as the relative portion of time spent
on $a$, and the number of jumps $b \rightarrow a$ minus the number of jumps $a\rightarrow b$
divided by $t$, respectively. If a trajectory is closed we have the current conservation
$\sum_{b\rightarrow a}J_{ba}({\bm\eta})=0$. Correspondingly, functionals $ u_V(\rho), w_{\bm A}(\bm
J)$ represent the time-average of function $V: M_0 \to \mathbb{R}$ relative to the distribution
$\rho$, and the weighted average of function $\bm A : M_1 \to S_1$ (the unit circle), with respect
to the distribution of jumps  $\bm J$.

The variational principle (\ref{variational-pinciple}) for the
discrete case can be derived exactly in the same fashion as for the
continuous case using the integration measures ${\cal
D}V=\prod_{a}dV_{a}$, and ${\cal D}{\bm A}=\prod_{a\rightarrow
b}(2\pi iz_{ab})^{-1}dz_{ab}$. The fields $V$ and $\bm A$ arise, as
in (\ref{S-modified}), from the representation of Dirac's
distribution for (\ref{prob}).  The integrations $dV_{a}$ and
$dz_{ab}$ go over the real axis and the unimodular circle
$|z_{ab}|=1$, respectively. As in the continuous case, the CDCF
${\cal S}({\bm J},\rho)$ can be computed using the Fokker-Plank (FP)
approach, which results in the expression (\ref{FP-modified}) with
the modified FP operator
\begin{align}
\label{FP-modified-discrete} 
 & \hat{{\cal L}}_{{\bm A},V}=
 \sum_{ab}k_{ab}z_{ab}\hat{\sigma}_{ab}-\sum_{a}\left(\kappa_{a}+V_{a}\right)\hat{\sigma}_{aa},
\end{align}
where $\hat{\sigma}_{ab}=|a\rangle\langle b|$. Note that for ${\bm J}, V=0$, the operator $\hat{{\cal
L}}$ represents the rate matrix of the master equation, so $\hat{{\cal L}}_{{\bm A},V}$ can
be referred to as the modified rate matrix.

Similar to the continuous case, the functional ${\cal S}({\bm J},\rho)$ can be identified
explicitly, by fixing the gauge using conditions $\psi_{{\bm A},V}=\rho_{a}$  and ${\cal F}({\bm
A},V)=0$. Thus, $\rho$ is the zero mode of the modified FP operator (\ref{FP-modified-discrete}).
This results in:

\begin{align}
\label{CDF-discerete} & {\cal S}({\bm J},\rho)=\sum_{ab}J_{ab}\zeta_{ab}+\sum_{a}\rho_{a}V_{a},
\\ &
\label{current-discrete-expl}
J_{ab}=k_{ab}z_{ab}\rho_{b}-k_{ba}z_{ba}\rho_{a}=k_{ab}z_{ab}\rho_{b}-k_{ba}z_{ab}^{-1}\rho_{a},
\end{align}
with $\zeta_{ab}=\ln(z_{ab})$. We further make use of
(\ref{current-discrete-expl}) to express $z_{ab}$ in terms of $({\bm
J},\rho)$ and of the gauge fixing condition ${\cal F}({\bm A},V)=0$
combined with the explicit form of the modified FP operator
(\ref{FP-modified-discrete}) to find an explicit expression for
$V_{a}$. Upon substitution of these expressions into
(\ref{CDF-discerete}) we arrive at the explicit expression for the
CDCF
\begin{align}
\label{CDDF-discerete-explicit} & {\cal S}({\bm
J},\rho)=\sum_{a}\kappa_{a}\rho_{a}-\sum_{ab}\sqrt{J_{ab}^{2}+4k_{ab}k_{ba}\rho_{a}\rho_{b}}
\nonumber
\\ & +\sum_{ab}J_{ab}
\ln\frac{\sqrt{J_{ab}^{2}+4k_{ab}k_{ba}\rho_{a}\rho_{b}}+J_{ab}}{2k_{ab}\rho_{b}}.
\end{align}
The variational principle can be applied to calculate the long-time distributions of the production
rates for the observables represented as linear functions of current/density. In particular we have
for the entropy production rate:
\begin{align}
\label{w-discrete} & w({\bm J})=\sum_{ab}J_{ab}\ln\left(k_{ab}/k_{ba}\right).
\end{align}

\paragraph{6.$\!$ Non-equilibrium thermodynamics for systems of identical particles}

In this section we illustrate how to derive a single-particle, coarse-grained, version of
(\ref{CDDF-continuous-explcit}). This is a potentially productive approach for addressing steady,
non-equilibrium situations in stochastic systems of interacting particles \cite{Sakaguchi,
Sherrington, CMS}.

Consider a system of  $N$ interacting Langevin processes,
\begin{eqnarray}
\dot{\eta}_i=F_i(\eta_1,\cdots,\eta_N)+\xi_i(t), \label{eta_multi} \quad \eta_i \in \mathbb{R}^1.
\end{eqnarray}
We introduce the multi-point density and currents depending on all  trajectories
$\{ \eta_i(\tau) \}_{i=1}^N $ and markers $\{ x_i \}_{i=1}^N$
\begin{eqnarray}
 & \varrho^{(N)}
  \equiv  \frac{1}{t}\int_0^t d \tau
  \prod_{k=1,\cdots N}\delta\left(\eta_k(\tau)-x_k\right),
 \la{rho-jay1N} \\
 & \gimel_i^{(N)}
 \equiv  \frac{1}{t}\int_0^t d \tau \dot{\eta}_i(\tau)
  \prod_{k=1,\cdots N}\delta\left(\eta_k(\tau)-x_k\right).
 \la{rho-jay2N}
\end{eqnarray}
The multi-point effective action (\ref{CDDF-continuous-explcit}) for the joint p.d.f. of the 
multi-point density and currents, 
$\langle\delta(\rho^{(N)}-\varrho^{(N)})\prod_{i=1}^N\delta(J_i^{(N)}-\gimel_i^{(N)})\rangle 
\propto \exp(-tS^{(N)})$, is
\begin{equation}
 \label{CDDF-continuous-explcitN}
 S^{(N)}\!\!=\sum_k\!\!\int \frac{d^N \!\!\bm x}{\rho^{(N)}}
 \left [F_k\rho^{(N)}-J_k^{(N)}
 -\frac{\partial_{x_k}\rho^{(N)}}{2}\right]^2.
\end{equation}
Averages over $i,\!$ $\varrho^{(0)} \equiv  \frac{1}{Nt}\int_0^t d \tau
 \sum_k\delta\left(\eta_k(\tau)-x\right)$,  $\gimel^{(0)} \equiv  \frac{1}{Nt}\int_0^t d \tau \sum_k\dot{\eta}_k(\tau)  \delta\left(\eta_k(\tau)-x\right)$ describe single-particle density/currents. The effective action that controls the long-time asymptotic of the single-point p.d.f. $\langle\delta(\rho^{(0)}-\varrho^{(0)})\delta(J^{(0)}-\gimel^{(0)})\rangle \propto \exp(-tS^{(0)})$ is related to the multi-point action (\ref{CDDF-continuous-explcitN}) via the  exact variational relation
\begin{eqnarray}
 && S^{(0)}=\left.{\mbox{inf}}_{\rho^{(N)},J^{(N)}_1,\cdots,J^{(N)}_N}
 S^{(N)}\right|_\Xi\label{S0}\\
 &&
 \Xi=\left\{\begin{array}{c}
 \rho^{(0)}(x)=N^{-1}\int\prod_i dx_i\sum_k\rho^{(N)}\delta(x_k-x)\\
 J^{(0)}(x)=N^{-1}\int\prod_i dx_i\sum_kJ_k^{(N)}\delta(x_k-x)
 \end{array}.\right.
 \nonumber
\end{eqnarray}
This can be derived by looking for the minimum of (\ref{CDDF-continuous-explcitN}), subject to constraints (\ref{S0}), using  Lagrange multipliers.

We now show  how to derive asymptotic results for this example of a many-body system of
interacting particles on a circle, with nearest-neighbor interactions, for a  {\emph{constant}} force field $F$. Starting from the discrete version of the model, we will consider the hydrodynamic limit $N \to \infty$ and obtain a smooth continuum model. Equations for interacting Langevin processes $\{ \phi_i \in S_1\}_{i=1}^N$  are
$$
\dot \phi_i = uN(\phi_i - \phi_{i-1}) + vN^2 (\phi_{i+1} + \phi_{i-1}-2\phi_i) + F + \xi_i(t),
$$
with $u, v$ effective parameters fixed by the interaction.
In order to consider the hydrodynamic $N\to\infty$ limit, we introduce the field $\varphi(i, t)$ whose
value at $i$ is the Langevin process $\phi_i$. A similar construction gives the noise field $\xi(i, t)$. In the
hydrodynamic limit, the indices form a continuum, $\frac{i}{N} \to y \in S_1$ and the evolution is given by
\be \label{abelian-Langevin}
\dot{\varphi}(y,t)=u\frac{\partial\varphi(y,t)}{\partial
y}+v\frac{\partial^{2}\varphi(y,t)}{\partial y^{2}}+F+\xi(y,t).
\ee
We can now apply the formalism presented earlier to compute the current  $J^{(0)}$ in the continuum limit. It will depend on the {\emph{connectivity component}}, $n$. This integer parameter allows to separate particles in groups according to the number of expected full cycles completed in the large time limit. To illustrate this, consider runners on a circular track, and separate the group which completes between $n$ and $n+1$ turns in a given time.
We call that group a {\emph {component}}, characterized by the integer $n$.

Performing a twisted Fourier decomposition
\be \la{twisted}
\varphi(y,t) = ny+
\varphi_{0}(t)+\sum_{k=1}^{\infty}\left[a_{k}(t)e^{iky}+b_{k}(t)e^{-iky}\right],
\ee
allows the minimization in (\ref{S0}) to be performed explicitly. This results in a set of Cr\'{a}mer
functions for current $J$ and density $\rho(\varphi)$, labeled by $n$
\begin{align}
\label{CDDF-continuous-reduced} {\cal S}_{n}^{(0)}(J, \rho)=N\oint \frac{d\varphi}{2 \rho}
\left [ (F+un)\rho-J- \frac{1}{2}\partial_{\varphi}\rho \right ]^2
\end{align}
By symmetry, minimization with respect to $\rho$ in (\ref{CDDF-continuous-reduced}) yields
$\rho(\varphi)=1$, and results in Gaussian distributions for $J$, peaked at $F+un$. The time-averaged current does not depend on the parameter $v$, which describes the {\emph{diffusive}} part of the process, but only on $u$, which is related to {\emph{drift}} around the circle. This is a natural conclusion upon analyzing averages of derivatives for (\ref{twisted}): except for the linear term $ny$, all others average to zero due to their oscillatory nature. Hence, contributions are expected only from the terms depending on first-order derivatives.

\paragraph{7. Notes on further potential applications} \label{notes}
Turbulence is often modeled via a set of stochastically driven
differential equations, which can be regularized in terms of
interacting Lagrangian particles subjected to stochastic stirring
and damping. Thus, a model describing the 3D evolution of a
representative turbulent blob in terms of four-point configurations
was presented in \cite{99CPS}. Considering coordinates and momenta
of the Lagrangian particles as degrees of freedom, one maps the
particle system onto the general model (\ref{bare}). Analyzing
various functionals of relevant densities and currents (introduced
for both the coordinates and momenta of the Lagrangian particles)
and relating them to standard objects of interest in stochastic
hydrodynamics, e.g., distribution functions of energy and density
fluxes in momentum space, constitutes an intriguing new direction
for future research.

\paragraph{8. Acknowledgements}
V.Y.C. acknowledges support through the start-up funds from Wayne State University. Research of
M.C. and R. T. was carried out under the auspices of the National Nuclear Security Administration
of the U.S. Department of Energy at Los Alamos National Laboratory under Contract No. DE
C52-06NA25396. V.Y.C. also acknowledges support from Los Alamos National Laboratory. M. C.
acknowledges support of the Weston Visiting Professorship program at Weizmann Institute.

\paragraph{Note} After the research described in this Letter was completed, we have learned about
similar results obtained recently in \cite{Maes}. Unlike in this work, they were obtained using a
standard formulation  \cite{BADZ}. We emphasize
that our approach is different and allows more  general types of interactions
between Langevin processes.

\newpage

\end{document}